# Enhancing narrowband high order harmonic generation by Fano resonances

Jan Rothhardt,[1,2] Steffen Hädrich,[1] Stefan Demmler,[1] Manuel Krebs,[1] Stephan Fritzsche,[2,3]
Jens Limpert,[1,2] and Andreas Tünnermann[1,2,4]

[1]*Institute of Applied Physics, Abbe Center of Photonics, Friedrich-Schiller University Jena, Albert-Einstein-Strasse 15, 07745 Jena, Germany*
[2]*Helmholtz-Institute Jena, Fröbelstieg 3, 07743 Jena, Germany*
[3]*Theoretisch-Physikalisches Institut, Friedrich-Schiller-Universität Jena, 07743 Jena, Germany*
[4]*Fraunhofer Institute of Applied Optics and Precision Engineering, Albert-Einstein-Strasse 7, 07745 Jena, Germany*
e-mail address: j.rothhardt@gsi.de

Resonances in the photo-absorption spectrum of the generating medium can modify the spectrum of high order harmonics. In particular, window-type Fano resonances can reduce photo-absorption within a narrow spectral region and, consequently, lead to an enhanced emission of high-order harmonics in absorption-limited generation conditions. For high harmonic generation in argon it is shown that the $3s3p^6np\,^1P_1$ window resonances (n=4,5,6) give rise to enhanced photon yield. In particular, the $3s3p^64p\,^1P_1$ resonance at 26.6 eV allows a relative enhancement up to a factor of 30 compared to the characteristic photon emission of the neighboring harmonic order. This enhanced, spectrally isolated and coherent photon emission line has a relative energy bandwidth of only $\Delta E/E=3\cdot 10^{-3}$. Therefore, it might be directly applied for precision spectroscopy or coherent diffractive imaging without the need of additional spectral filtering. The presented mechanism can be employed for tailoring and controlling the high harmonic emission of manifold target materials.

**Keywords:** High harmonic generation, Fano resonances

**PACS:** 32.30.-r, 42.65.Ky, 42.50.Hz

High harmonic generation (HHG) driven by a strong laser field represents an attractive method for generating coherent radiation in the extreme ultraviolet spectral region [1,2] and is, nowadays, widely employed in atomic, molecular, plasma and solid state physics. Recently, even keV photon energies have been demonstrated by HHG providing a bright source of coherent X-rays [3]. In addition, HHG allows producing extremely short attosecond pulses, which enable groundbreaking investigations in the field of attosecond science [4]. Typically, HHG suffers however from its inherently low conversion efficiency, which hinders applications that are especially dependent on a high photon flux in order to achieve a good signal-to-noise ratio.

In a simple single atom picture HHG can be described by a three-step model [5]: First, an electron tunnel-ionizes through the atomic potential, which is modified by the strong laser field. Second, the electron propagates in the strong laser field. Finally, the electron may recombine with its parent ion and gives, thus, rise to the emission of a photon (third step). In this picture, the power emitted by a single atom at the frequency $\omega_q$ is proportional to the square of the amplitude of the oscillating dipole $|A_q(t)|^2$, which is induced by the recombination process [5].

While the response of a single atom to the laser field ($A_q$) is extremely small, phase matching of a large number of emitters gives rise to a coherent build-up of high-order harmonic (HH) emission along the propagation direction of the driving laser through the generation medium.

By optimizing the macroscopic generation conditions, hence, maximizing the number of phase-matched emitters, conversion efficiencies as high as of $4\cdot 10^{-4}$ at 73 nm (17 eV) [6] and $5\cdot 10^{-5}$ at 53 nm (23.3 eV) [7] have been achieved with 800 nm driving wavelength. With the help of a simple model, which takes the emission of all emitters along the propagation axis, phase-mismatch and absorption into account, Constant et al. [7] found that the maximum conversion efficiency, which can be achieved by HHG, is ultimately limited by reabsorption of the generated high harmonic photons within the generating medium itself. Furthermore, it was shown in Ref. [7] that this so-called absorption-limited conversion efficiency is proportional to $|A_q/\sigma_q|^2$, with $\sigma_q$ being the absorption cross-section at the harmonic frequency $\omega_q$. Thus, the absorption-limited conversion efficiency of HHG can only be increased, by increasing the emission due to the dipole amplitude ($A_q$) or by reducing the re-absorption ($\sigma_q$) of HH photons.

So far large efforts have been devoted to increasing $A_q$. For example molecules [8] and noble gas clusters [9–11] have been employed as generating medium.

Resonances involved in the generation process have been also explored as a promising route towards higher conversion efficiency by increasing the single-atom response $A_q$ [12–14]. In particular, Ganeev et al. observed a significant resonant enhancement of a single harmonic order in laser-generated indium plasma plumes [15,16] which they attribute to resonant transitions between autoionizing states and the ground state of singly charged transition metal ions. Their work inspired a number of theoretical and experimental studies about the physical nature of the enhancement process [17–20]. Recently, Strelkov et al. [18] suggested the following extension of the recombination (third) step of the established three-step model of HHG [5] in order to explain the experimentally observed enhancement: Instead of radiative recombination from continuum to the ground state, the free electron can also be trapped into an auto-ionizing state followed by a transition to the ground state and photon emission. In the particular case that recombination via the autoionizing state has a higher probability than direct recombination, an enhancement of HHG at the specific transition energy can be observed.

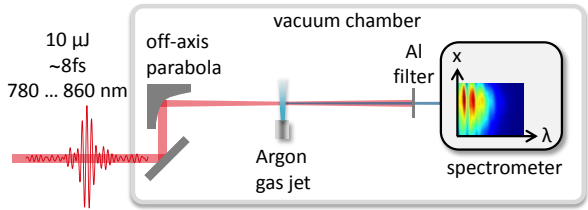

FIG. 1. (Color online) Experimental setup for high harmonic generation: A few-cycle laser pulse, whose central wavelength is tunable between 780 and 860 nm, is focused by an off-axis parabola upon an argon gas jet, which emerges from a 150 μm diameter nozzle. A 200μm thick aluminum filter separates the fundamental laser radiation from the generated XUV photons, which are analyzed spectrally and spatially by means of a grating-based spectrometer.

Here, we report on the first experimental observation of resonant enhancement of high-order harmonics in argon. In particular, we analyze the underlying physical mechanism experimentally and suggest a simple model for its explanation. We conclude that in our case the resonant enhancement of the macroscopic HH emission is dominated by reduced absorption due to Fano-resonances in the photo-absorption spectrum of the generation medium. Notably, the enhancement mechanism is of different nature than what has previously been observed for various transition-metal ions [15,16].

The experiments are carried out using a few-cycle optical parametric chirped pulse amplifier (OPCPA) as driving laser [21]. Owing to the large bandwidth supported by the OPCPA system, the central wavelength of the laser can be tuned between 780 nm and 860 nm, with negligible changes in pulse duration (~8 fs), pulse energy (~10 μJ) and beam parameters. The experimental setup is shown in Fig.1. An off-axis parabolic mirror is employed for focusing the laser pulses to a spot size of $2w_0=30$ μm, which enables a peak intensities of $\sim 2\cdot 10^{14}$ W/cm$^2$. The target gas is provided by a 150 μm diameter nozzle. By placing the gas jet slightly behind the focus of the driving laser, phase matching of the short trajectories has been achieved and optimized for the highest on-axis harmonic emission [22]. The fundamental radiation of the driving laser is blocked by a 200 nm thick aluminum filter, while the generated XUV radiation is analyzed spatially and spectrally by means of a grating spectrometer incorporating a gold-coated variable line spacing flat-field grating (nominal 1200 lines/mm) and a XUV-sensitive charge-coupled device (CCD) camera. Since only the spatial dimension perpendicular to the grating lines is focused to the detector plane, the XUV beam propagated freely in the other spatial dimension enabling spectrally resolved one-dimensional beam profiles to be recorded. The energy resolution of the spectrometer has been measured to be better than $\Delta E/E=2\cdot 10^{-3}$ using narrow bandwidth radiation.

Figure 2 displays a series of spectra that have been recorded with backing pressures between 1 bar and 9 bar of argon and the central wavelength of the driving laser tuned to~820 nm. At a backing pressure of 1 bar (black) four distinct harmonic orders H15, H17, H19 and H21 are clearly observed. Due to the short pulse duration of our driving laser the individual harmonic orders appear as rather broad lines in the spectrum. At higher backing pressures the two harmonics H15 and H17 are gradually suppressed. In addition, various narrow lines appear in the spectrum between 26 eV and 30 eV. A particularly strong and narrow line (FWHM~0.1 eV, $\Delta E/E=3\cdot 10^{-3}$) is observed at 26.6 eV and 9 bar backing pressure (pink). This line is shown on a logarithmic scale (pink) together with the underlying harmonic H17 in fig. 2 b). Note that the peak intensity of this enhanced line is a factor of 30 higher than the peak of the underlying harmonic H17 which is represented by a Gaussian fit (black).

Figure 2 c) shows the photo-ionization (PI) cross-section that has been measured by means of synchrotron radiation [23]. Since the photon energy of interest here is well above the ionization threshold, the PI cross-section dominates the photo–absorption (PA) and represents the corresponding cross-section very well. The strong suppression of the low order harmonics (H15 and H17) at high backing pressures can be explained by the dramatic increase of the PA cross-section from only 17 Mbarn at H21 to 35 Mbarn at H15

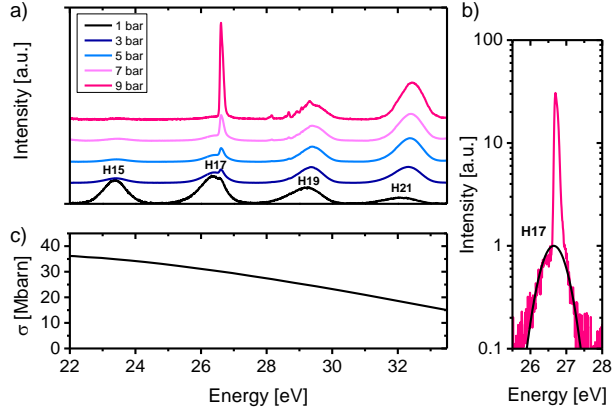
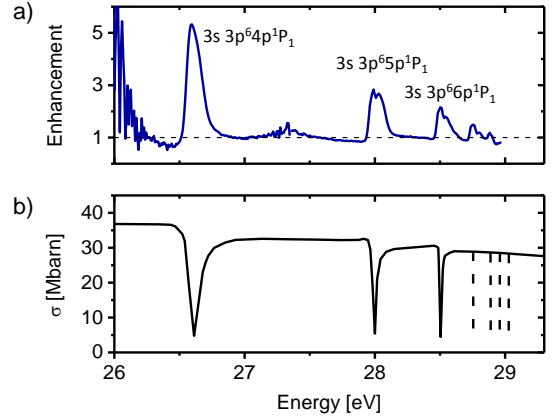

FIG. 2. (Color online). **a)** Spectra of high order harmonics generated in Argon at different backing pressures between 1 bar and 9 bar. Each spectrum is normalized to the peak of H19 (at 29.2 eV). At low pressure (black line) four harmonic orders (H15 to H21) are detected. While the four harmonic orders H15 to H21 are clearly observable at 1 bar (black line), H15 and H17 are gradually suppressed due to absorption as the pressure increases. Instead, a narrow but rather intense line occurs at 26.6 eV on top of harmonic H17 and a number of similar but weaker lines are observed between 29 eV and 30 eV. **b)** Logarithmic plot of the spectrum around the strongest enhanced line at 26.6 eV measured at 9 bar (pink line) normalized to the peak of H17 (black line). **c)** Measured continuous photo-ionization cross-section reproduced from [23], which accounts for direct photoionization only without including any resonances.

FIG. 3. (Color online). **a)** Spectral profiles of the observed resonances extracted from measurements of high harmonic spectra generated at 7 bar backing pressure. Details on the calculation of the enhancement factor can be found in the text. Each enhanced line is attributed to a transition of an autoionizing-state (named in the figure) to the ground state ($3S^23p^6$). Line splitting is observed for the $3s3p^65p^1P_1$-$3S^23p^6$ and $3s3p^66p^1P_1$-$3S^23p^6$ transition. **b)** Photoionization cross-section including the $3s3p^6np^1P_1$ series of window-type resonances measured by Madden et al. [24]. The dashed line indicate the position of the non-resolved resonances with n>6.

The origin of the narrow lines that are observed at high backing pressure refers to a series of resonances involved in photo-ionization of argon [24]. The enhanced lines observed in the high order harmonic spectra (Fig. 2. a)) clearly belong to the Fano-resonances that are well-known from photo-ionization spectra [24]. These resonances can be understood as interplay between the direct ionization (amplitude) and the photo-excitation of one or two electrons to some valence or Rydberg state, followed by a subsequent autoionization process. Because of the different phases that are associated with these 'quantum paths' (quantum amplitudes), the various shapes of isolated resonances can be described in terms of Fano profiles [25]. In the case of Argon the autoionizing states of the $3s3p^6np^1P_1$ (n=4,5,6) series result in so-called "window resonances", for which the direct and autoionization amplitudes virtually cancel each other and, hence, the cross-section of photo-ionization is dramatically reduced. These Fano-resonance can be clearly seen in the PI cross-sections, which have been measured by Madden et al. [24] and displayed in Fig. 3 b).

The measured HH spectra allow extracting the spectral line profiles of the observed resonances. To this end, an enhancement ratio has been introduced and is calculated as the photon signal in presence of the resonance divided by the expected signal without the resonance, which is found by a Gaussian fit to the underlying harmonic.

The signal-to-noise ratio of this method has been maximized by tuning the peak of either H17 or H19 close to the resonance of interest via changing of the driving laser wavelength. Figure 3 a) displays the extracted line profiles on top of the measured PI cross-sections including the Fano-resonance (Fig. 3 b)).

A number of experimental investigations have been carried out in order to prove that the observed enhanced radiation is coherent, has properties similar to the high order harmonics and does not simply refer to spontaneous atomic line or plasma emission. First of all, the beam profile (Gaussian-like) and the divergence (half angle ~5.4 mrad) of the enhanced emission and the underlying harmonic are found to be identical. This proves directional emission, coupled to the wave front of the driving laser. Secondly, at low backing pressures a quadratic increase of the signal with backing pressure is observed for both the enhanced lines and the neighboring harmonics indicating coherent emission of the increasing number of emitters.

This pressure dependence is illustrated in Fig. 4. The dots represent the measured power at the center of H15 (dark blue), H17 (light blue), H19 (pink) and the line enhanced by the $3s3p^64p^1P_1$ resonance (black) versus backing pressure. It can be seen that H15, H17 and H19 peak at about 3.5 bar, 4 bar and 4.5 bar, respectively. However, the enhanced line increases until 5 bar of backing pressure and decreases much slower than the harmonics when the backing pressure is further increased.

To investigate the effects of different backing pressure on phase matching and absorption we employ the model introduced by Constant et al. [7]. The number of photons emitted on axis into the q-th harmonic per unit of time and area can be calculated as [7,26]:

$$N_{out} \sim \rho^2 A_q^2 \frac{4L_{abs}^2}{1+4\pi^2\left(L_{abs}^2/L_{coh}^2\right)} \cdot \left[1+\exp\left(-\frac{L_{med}}{L_{abs}}\right)-2\cos\left(\pi\frac{L_{med}}{L_{abs}}\right)\exp\left(-\frac{L_{med}}{2L_{abs}}\right)\right], \quad (1)$$

when $A_q$ and the gas density $\rho$ are assumed to be constant within the generating volume. The pressure dependent absorption length $l_{abs}=1/\rho\sigma_q$ is calculated via the measured continuous photo-ionization cross-sections of argon [23]. The target pressure p and, consequently, the density $\rho$ in the interaction region is calculated to be about 40% of the backing pressure, if we assume that the center of the laser beam is as close as 30 μm ($2w_0$) to the nozzle opening [27]. The coherence length $l_{coh}=\pi/\Delta k$ is determined by the phase mismatch $\Delta k$, which is calculated as the difference between a pressure-independent Gouy phase term $\Delta k_{Gouy}$ and the pressure dependent dispersion of the gas atoms $\Delta k_{Disp}=(p/p_0)\cdot(2\pi/\lambda_0)\cdot\Delta\delta$. Here $p_0$ is the standard pressure, $\lambda_0$ is the driving laser wavelength and $\Delta\delta$ is the refractive index difference between fundamental and harmonic taken from [28]. Negligible ionization and negligible dipole phase [22] is assumed and additional absorption due to residual gas in the vacuum chamber is taken into account.

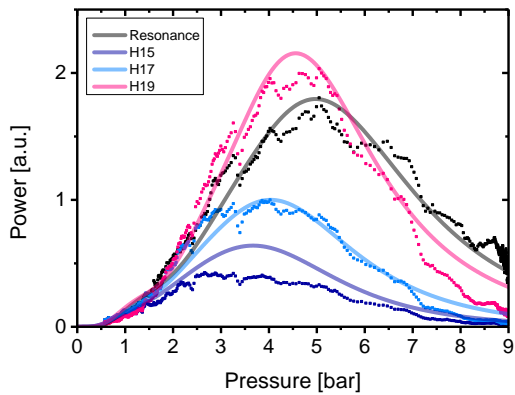

FIG. 4. (Color online). Measured power at the center of H15 (dark blue dots), H17 (light blue dots) and H19 (pink dots) versus the applied argon backing pressure. The measured power at the center of the strongest resonant line (26.6 eV) involving the $3s3p^64p^1P_1$ state is shown as black dots. All values are normalized to the maximum power obtained for H17. The semitransparent lines represent the result of a corresponding simulation accounting for phase matching and absorption effects. Details about the simulation and the used parameters can be found in the text. Due to the characteristics of the utilized vacuum pump the density of residual gas can be assumed to be proportional to the backing pressure. The constant phase mismatch term $\Delta k_{Gouy}$ and the length pressure product of the additional absorbing atoms have been used as fit parameters to best match the experiment. The results of the simulation are displayed as solid lines in fig. 4 (dark blue, light blue and pink line) and agree well with the measurements. The pressure dependence of the resonant line at 26.6 eV can be reproduced by the model as well (black line), if the dispersion ($\Delta\delta$) is reduced by 15%. Furthermore, the effective absorption cross-section is determined to be 25+/- 1 Mbarn, which is significantly lower compared to the background cross-section of 30 Mbarn that has been measured at this energy [23]. Hence, the model reproduces the experiments very well and strongly supports our interpretation of reduced absorption due to the resonance.

In a second step, the model is employed to investigate the spectral characteristics of the resonance.

The high harmonic orders are well reproduced by representing each harmonic order in $A_q(\omega)$ with a Gaussian profile of adequate width. In addition, the absorption cross-section was modeled by including the strongest resonance using Fano's parameterization [25]. Here the resonance energy and width as well as the q-parameter and correlation index (that describe the shape and the strength of the resonance) have been used as fit parameters. The simulated emission spectrum was corrected by the known spectrometer resolution (black dashed line in Fig. 5 a)) and agrees remarkably well with the measured spectrum (blue line in Fig. 5 a)). The corresponding effective absorption cross-section is shown in Fig. 5 b). The effective half-width of the $3s3p^64p^1P_1$ resonance in our experiment is found to be $\Gamma=0.18$ eV, the correlation index is $\rho^2=0.175$ and the q-factor is q=-0.55. These values differ from what has been measured in photo-ionization with synchrotron radiation ($\Gamma=0.08$ eV, $\rho^2=0.86$ and q=-0.22) [24] but the differences can be easily attributed to the fact that the resonances are influenced by the strong laser field. While time-resolved absorption measurements have already unveiled a laser-induced spectral shift, weakening and line splitting of the involved resonances [29], here we observe a time-averaged response of the absorbing medium resulting in broadened resonance lines.

The presented mechanism is universal and is expected to be observable in other target materials as well. Indeed, our setup allows observing a significant enhancement of a narrow spectral line due to the $4s4p^65p^1P_1$ window resonance at 25 eV in Krypton. Furthermore, we observed suppression of three narrow spectral lines between 45 eV and 47.2 eV in Neon, due to resonances involving the $2p^43s3p^1P_1$, $2s2p^63p^1P_1$ or $2s2p^64p^1P_1$ states, which lead to increased absorption [30]. Similar resonances involving autoionizing states have been faintly observed by HHG in Helium [31], are known for Xenon [32] and can be found in many other multi-electron systems.

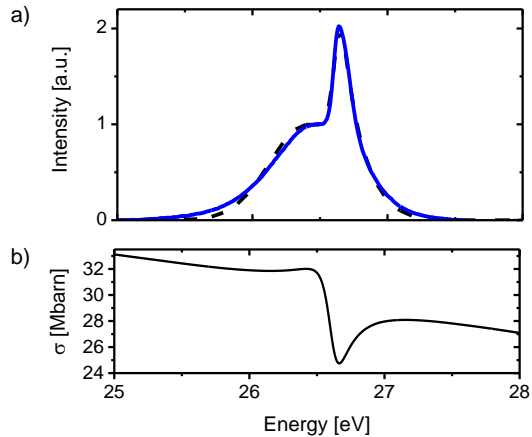

FIG. 5. (Color online). Zoom into the spectral region containing only H17 and the resonance caused by the $3s3p^6 4p$-$3S^2 3p^6$ transition at 26.6 eV. **a)** Measured spectrum at 5 bar backing pressure (blue line) and spectrum calculated from a corresponding simulation (black dashed line). **b)** Absorption cross-section used for calculating the spectrum presented on top (black dashed line fig. 5 a)). A Fano resonance profile has been added to the measured direct photoionization cross-section (Fig. 2b) taken from [23] and the effective width, q-factor and correlation index of this resonance have been chosen to match the experimental result best.

In summary, we demonstrate that resonances in the photo-absorption spectrum of the generating medium can significantly affect and modify the spectrum of high order harmonics under absorption-limited conditions. First experiments in argon show that window resonances due to the virtual excitation of autoionizing states, give rise to enhanced photon emission. In particular, the $3s3p^6 4p^1 P_1$ resonance at 26.6 eV allows a relative enhancement by a factor of 30 compared to the neighboring harmonic order. This enhanced line is spectrally isolated and possesses a relative energy bandwidth of only $\Delta E/E = 3\cdot 10^{-3}$ without the need for additional spectral filtering. Therefore, it is very attractive for directly driving a huge variety of applications requiring narrowband XUV radiation.

The presented mechanism provides a simple and easy to implement way of tailoring and enhancing high harmonic emission and can be applied to a huge variety of targets. In combination with high average power driving lasers [33] it will enable table-top XUV sources of unprecedented brightness and spectral purity. Thus, applications such as precision spectroscopy of highly charged ions [34] or coherent diffractive imaging of nanoscale objects with unprecedented resolution and level of detail [35] will be feasible with table-top narrowband HHG sources in future.

This work has been supported by the German Federal Ministry of Education and Research (BMBF) and the European Research Council (FP7/2007-2013)/ERC Grant agreement no [240460]. We thank Pascal Salières for fruitful discussions.